# Dynamic MPLS with Feedback

Ankur Dumka[1] and Prof. Hadwari Lal Mandoria[2]

[1]Ph.D Scholor , Uttarakhand Technical University, INDIA

ankurdumka2@gmail.com

[2]College of Technology, G.B.Pant University of agriculture and technology, Pantnagar, INDIA

drmandoria@gmail.com

## ABSTRACT –

*Multiprotocol Label Switching (MPLS) fasten the speed of packet forwarding  by forwarding the packets based on labels and reduces the use of routing table look up from all routers to label  edge routers(LER) , where as the label switch routers (LSRs) uses Label Distribution Protocol (LDP) or RSVP (Resource reservation Protocol) for label allocation and Label table for packet forwarding .*

*Dynamic protocol is implemented which carries a Updates packets for the details of Label Switch Paths, along with this feedback mechanism is also introduced which find the shortest path among MPLS network and also feedback is provided which also help to overcome congestion, this feedback mechanism is on a hop by hop basis rather than end to end thus providing a more reliable and much faster and congestion free path for the packets .*

## 1.INTRODUCTION

Multiprotocol Label switching (MPLS) is an IETF standard that merges layer 2 and layer 3 protocol that uses label switching in the core network, thus reduces the workout of looking the routing table overhead. MPLS is a traffic switching technology that combines the traffic engineering capability of Asynchronous Transfer Mode (ATM) with flexibility and scalability of Internet Protocol (IP). This enables fast packet transfer over the network. MPLS establishes a connection-oriented mechanism into a connectionless IP network. MPLS uses short, length-fixed, locally significant labels in the packet header between layer 2 and layer 3 header and the packets are forwarded according to label rather than by routing protocol in the core of network. Egress routers use the routing table to convert IP packet into MPLS label and back to IP packet while leaving the network. MPLS technology offer services including layer 3 Virtual Private Network (VPN) virtual Private Remote Network (VPRN), traffic engineering (TE), traffic protection and layer 2 VPN Virtual Path Lan service(VPLS).

In Multiprotocol Label Switching (MPLS) a table is created in the control plane for forwarding of packets , there are two tables in the control plane that are routing table known as routing information base(RIB) and label table called Label Information Table (LIB). These information is used to find the path for shortest route in the MPLS network . These paths are used for forwarding of packets. These paths are forwarded to data plane where a table is created for routers and labels  known as forwarding information base (FIB) and Label Forwarding information base (LFIB) which contains path selected by control plane for forwarding of labels and packets.

In control plane, Label Distribution Protocol (LDP) is used in the control plane for  finding paths for the labels within the MPLS network . Resource Reservation(RSVP) protocol is used for Traffic Engineering (TE). There are routers at the edge of the MPLS network are called as Label Edge Routers (LER) and other routers are known as Label Switch Routers (LSR). As the packets enters the MPLS network into the Label Edge Routers, the packets are transferred to the next hop by seeing RIB , then after crossing LER and enter into LSR , routers look LIB and forward packets based on labels rather then routing table, as the label table contains information of routers in MPLS network only hence its size is smaller than routing table hence it enhances and make packet forwarding faster.Packets enter the MPLS network from the Ingress LER and make an exit from MPLS network from Egress LER. Ingress LER uses routing table and label table to change the packets into label and attaches a 32 bit header of MPLS label . Now, LSRs only uses their label table for swapping of label information for forwarding of packets to next hop. Same way the Egress LER changes labels into packet based forwarding by matching among routing table and label table.

The rest of this paper is organized as follows. Section 2 gives overview of MPLS, Section 3 describes the method and methodology used for the design Section 4 compares the algorithm with predefined algorithm and prove this as better than others Section 5 describe the Result and paper concluded in section 6 .

## 2. OVERVIEW

Update packets are passes from the edges routers to the network which gives the details of the network. There are two update packets are used in this network . Route update (RU) is used to show all the paths of the network . Best Path (BP) will provide the best path among them and also a backup paths for the destination . Now , as the packet enters in the MPLS network RU and BP will give a path from source to destination and also the best and backup path to the destination. Now, LDP is used to encapsulate a label in front of packet to transfer a packet to the destination . Now if all the packets forward to the destination with the smallest path then this path become congested and a feedback is send to source router by attaching a header of 1 bit to the packet from the alternate LSP which gives the status of the path. If this path become congested then value of this extra added bit will be 1 and if the path is free for other packets that is congestion free path then the value of this extra bit will be 0. On getting the value of 1 of this extra bit , the traffic is diverted to next backup path and Updates packets then again send and best path and backu  path selected which gives a dynamic and more relaible method for traffic and congested network.

# 3. METHOD AND METHODOLOGY

There are routers at the edges of MPLS enabled routers known as Edges routers , the routers at the incoming edge are called ingress edge router and router at egress side are called egress edge router, the label is attached at the ingress edge router and label removed at egress edge routers. The labels get swapped at the intermediate routers known as switching routers. Now RU packet is send from the routers which gives the detail of path in the network at regular intervals thus the path gets created and changes dynamically. Whereas, BP will select a best path and backup (second best) path among the path generated by LU and update the label table according to this two path only, thus extend the functionality of forwarding table by providing two path a best path and a backup path. Now, as the packet enters in the MPLS network, the ingress edge router based on the path selected by LU attach a label selected by LDP protocol. Now as the table is updated according to the best and backup path, so the switching routers switch the label based on the Label forwarding table (LFIB) and the packets keep on forwarding until reaches to the egress edge router. At the egress edge router the label is removed and the packet is forwarded based on packets IP address.

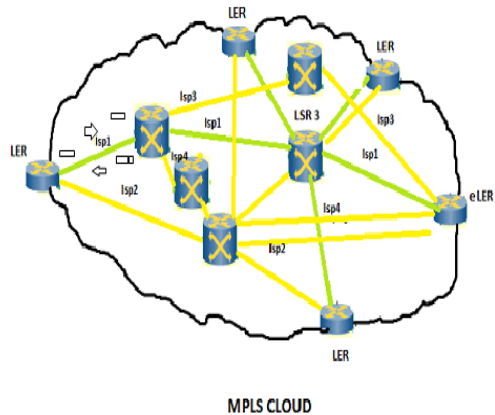

Fig. 1  MPLS network

Each routers contain two tables Routing Information Base (RIB) and Label Information Base (LIB) in the control plane and two table  Forwarding Information Base (FIB) and Label Information Base (LFIB) in the data plane. RIB contain information of routers and LIB contain information of label, whereas FIB contain best routes of routers and LFIB contain label information related to the best routes.

Feedback mechanism is indroduced in this way.The packet transferred from LSP 1 from source router to destination router and coming back from destination to source router by alternate path as LSP is unidirectional so a bit is added to the packet coming back from LSP reaches to router from alternate LSP which provide a feedback mechanism in the network and the value of this bit defined the congestion in the network.

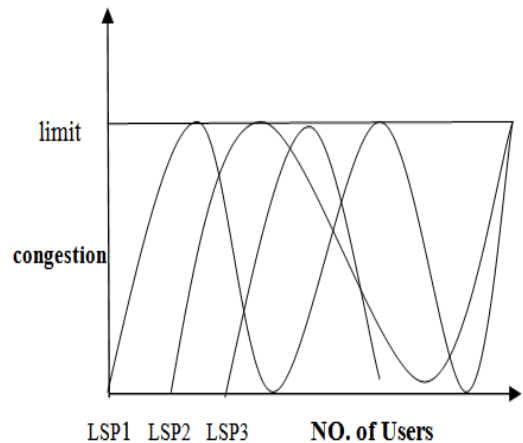

Fig. 2  Switching of LSP

# 4. BENCHMARK ALGORITHM USED FOR COMPARISON

For evaluating performance of our proposed algorithm, we have selected four algorithm, which we considered as benchmark for comparison:

1.) Optimal congestion with N+1 Label
2.) Random Races (RR) [5]
3.) Shortest Widest Path (SWP)
4.) Widest Shortest Path (WSP)

**Optimal Congestion with N+1 Label** -  In this algorithm an extra bit is attached and feedback mechanism is introduced in case of congestion occur in the network. The numbering of path is done statically as per the number of hops or djakastra's algorithm. Since all the packets travel from the shortest path hence a feedback is introduced which is included as adding a bit to the packet end whose value will determine the congestion in the network and in case of congestion forward that packet to second path.

**Random Races (RR) -** This algorithm proposed by Oomen and Mishra. This proposed and algorithm of numbering of paths and forwarding of packets based on this numbering. Traffic engineering and traffic management can also be done in this algorithm based on requirement and need of service provider in specific route.

**Shortest Widest Path (SWP) -** This algorithm is proposed by Wang and Crowcroft. Here there are two variants of algorithm distance vector based SWP and link state based SWP, in both case, algorithm find route with widest path that have maximum bandwidth. If there is tie then algorithm choses path having minimum length and minimum propogation delay. If there is again tie then any one path is randomly chosen.

**Widest Shortest Path (WSP) -**This algorithm proposed by Guerin et al. This algorithm chooses shortest path as first priority and in case of tie it choses maximum bandwidth or widest path and in case of tie in both cases a random path is chosen.

Now our algorithm is better than all the above in the case that apart from selecting the best route from the given routes, it chooses the path in a dynamic manner, and also having a feedback mechanism on a hop by hop mechanism and backup path in case of congestion . Thus comparing with all the above algorithm proposed algorithm prove to be a better than other algorithm.

# 5. RESULT

From the above given algorithm it is clear that LFIB provide us two best path one serve as primary path and other as backup and when congestion occur in the primary path that is detected by ATCC device[7] then this congestion can be detected and overcome by feedback mechanism . Thus this algorithm create a backup or alternative path for the same source and destination. Thus, in case of congestion a 1 bit label is attached at the end of MPLS label which designate the congestion in the network[3] and this label is attached to the packet coming back to the source router from alternate LSP, it will forward a packets from alternate path and a new alternate path is chosen. The benefit of this algorithm is that there is a short table lookup which reduces time for table lookup and also a backup path which handles the congestion thus provides a congestion free path even in case of increased number of users.

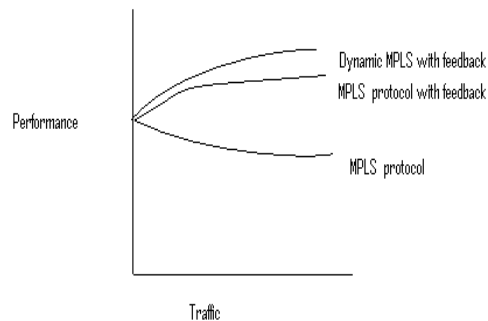

Fig. 3 Performance comparison of dynamic MPLS with feedback protocol with other protocols

# 6. CONCLUSION

This paper proposed a methodology of providing the congestion free path in a network even with increase in number of users within a network or increase in the traffic flow in the network and this was achieved by providing a feedback mechanism or congestion and error control mechanism to hop by hop or per router basis and also providing a backup path with dynamic protocol used to find the best path. Thus proposed algorithm selects a dynamic method of selecting best path by using LU and BP which select a backup path within MPLS network. A feedback mechanism is introduced for detecting a congestion within a network Congestion will be under the limit and will not exceed that limit inspite of increase in number of users in the network.

As future work, we try to implement services in this network using this algorithm and test it on real network.

# ACKNOWLEDGEMENT

I would say a special thanks to my supervisor and guide Dr. Hardwari Lal Mandoria  for helping me and